\documentclass[11pt, a4paper]{article}
\usepackage{fullpage}

\usepackage{belov_paper}
\newcommand{\maps}[1]{\;\stackrel{#1}{\longmapsto}\;}
\newcommand{\transduce}[1]{\stackrel{#1}{\rightsquigarrow}}

\def\inputcolor{red}
\def\markedcolor{blue}

\title{Global Phase Helps in Quantum Search: Yet Another Look at the Welded Tree Problem}
\author{
  Aleksandrs Belovs\thanks{Faculty of Computing, University of Latvia}
}
\date{}

\usepackage[framemethod=TikZ]{mdframed}
\usepackage{chngcntr}

\newmdenv[%
  roundcorner=5pt,
  linecolor=blue!15,
  linewidth=2pt,
  subtitlebackgroundcolor=blue!15,
  subtitleaboveskip=0pt,
  subtitlebelowskip=0pt,
  subtitleinneraboveskip=0pt,
  innerbottommargin=0pt,
  subtitlefont=\normalfont
]{mdfigure}

\def\myfigureInternal#1#2#3{
\refstepcounter{figure} #1
\begin{mdfigure}[
  frametitle={
    \tikz[baseline=(current bounding box.east),outer sep=0pt]
    \node[anchor=east,rectangle,fill=blue!15,rounded corners]
    {Figure~\thefigure};},
  frametitleaboveskip=-10pt
  innertopmargin=0pt,
  innerbottommargin=0pt,
  roundcorner=5pt,
  linecolor=blue!15,
  linewidth=2pt,
  subtitlebackgroundcolor=blue!15,
  subtitleaboveskip=0pt,
  subtitlebelowskip=0pt,
  subtitleinneraboveskip=0pt,
  subtitlefont=\normalfont
]
#3
\mdfsubtitle{\medskip #2}
\end{mdfigure}
}

\def\myfigure#1#2#3{
\begin{figure}[htb]
\myfigureInternal{#1}{#2}{#3}
\negbigskip
\end{figure}
}

\PassOptionsToPackage{svgnames}{xcolor}
\usepackage{tikz}
\usetikzlibrary{graphs, shapes.geometric, arrows.meta, matrix}
\tikzset{>={Stealth[width=2mm,length=2mm]}}

\begin{document}
\maketitle

\begin{abstract}
Up to now, relatively few exponential quantum speed-ups have been achieved.
Out of them, the welded tree problem~\cite{childs:walkExponentialSeparation} is one of the unusual examples, as the exponential speed-up is attained by a quantum walk.
In this paper, we give a very short proof of the optimal linear hitting time for this problem by a discrete-time quantum walk, which is based on a simple modification of the electric quantum walk framework~\cite{belovs:electicityQuantumWalks}.
The same technique can be applied to other 1-dimensional hierarchical graphs, yielding results similar to~\cite{balasubramanian:hierarchicalGraphs}.
\end{abstract}

\section{Introduction}

The usual behaviour of a random walk is that it mixes to the steady state.
This is caused by the exponential decay of the coefficients of the eigenvectors with small eigenvalues.
Quantum walks, on the other hand, are unitary operators, and they preserve the weight of its entire spectrum.
This leads to the following interesting phenomenon.
If the graph is carefully crafted, a quantum walk initiated at one specific ``focal'' vertex can efficiently traverse the whole graph -- even if it is huge -- and concentrate in some other ``focal'' vertex.
%This happens even if the graph is very large, and indeed results in exponential quantum speed-ups.

In early 2000s, this observation was used by Kempe~\cite{kempe:walkHitFaster} and Childs, Farhi, and Gutmann~\cite{childs:gluedTrees} to construct quantum walks exponentially outperforming their random counterparts.
Kempe~\cite{kempe:walkHitFaster} used a Boolean hypercube with the focal vertices being its two opposite vertices.
Childs~\etal~\cite{childs:gluedTrees} used two complete binary trees of depth $n$ glued at the leaves with the focal vertices being the roots of the trees.

Although the above two papers demonstrate an exponential separation between quantum and random \emph{walks}, they do not establish a similar separation between quantum and random \emph{algorithms}, as the corresponding search problems can be efficiently solved even by a deterministic algorithm~\cite{childs:walkExponentialSeparation}.
This lead Childs, Cleve, Deotto, Farhi, Gutmann, and Spielman~\cite{childs:walkExponentialSeparation} to the formulation of the welded tree problem.
Unlike glued trees, the leaves of the two complete binary trees in the welded tree problem are joined by a random cycle (see \rf{sec:weldedPrelim} for more detail).
Given a root of one of the trees, the task is to find the root of the other tree.
The quantum walk still traverses this graph with ease, while no randomised algorithm can solve this problem in sub-exponential time.
This results in an exponential quantum speed-up.

Childs at el.~\cite{childs:walkExponentialSeparation} used a \emph{continuous-time} quantum walk, and showed that its execution for time $O(n^{9/2})$ is sufficient to find the root of the opposite tree, where $n$ is the depth of the tree.
It was later improved to $O(n^2\log n)$ by Atia and Chakraborty~\cite{atia:improvedQuantumWalks}.
Li, Li, and Luo~\cite{li:weldedTreees} proved that analogous \emph{discrete-time} quantum walk requires $O(n^{3/2})$ iterations for the same task.
Balasubramanian, Li, and Harrow~\cite{balasubramanian:hierarchicalGraphs} generalised the continuous-time quantum walk of~\cite{childs:walkExponentialSeparation} to other types of hierarchical graphs.
While the algorithms in all these papers are based on standard quantum walks, rather elaborate spectral analysis is required to prove their correctness.

Discrete-time quantum walks are generally better suited for implementation in the traditional quantum circuit model as they avoid a complicated task of simulating a Hamiltonian.
Also, in many cases, they avoid spectral analysis.
In a seminal paper, Szegedy~\cite{szegedy:walk} showed that a discrete-time quantum walk attains a quadratic speed-up for detecting presence of marked vertices when initiated in the steady state of the random walk.
Electric quantum walk by Belovs~\cite{belovs:electicityQuantumWalks} is a generalisation of this result for arbitrary initial distribution.
Applicable to any bipartite graph $G$, electric quantum walk can detect presence of marked vertices in $O(\sqrt{WR})$ iterations, where $W$ is the total weight of the graph, and $R$ is the effective resistance between the initial and the marked vertices (see \rf{sec:walk} for definitions).
It is relatively easy to estimate both quantities, which allows development of quantum algorithms by purely combinatorial means~\cite{belovs:learning, lee:learningTriangle, belovs:mergedWalk3Dist}.
A big drawback of this approach is that $O(WR)$ captures hitting time of a \emph{random} walk, which means that electric quantum walks do not provide more than a quadratic speed-up.

This quadratic barrier was recently broken by Jeffery and Zur~\cite{jeffery:kDist} in a very interesting way.
They extended electric quantum walks with alternative neighbourhoods of vertices.
The algorithm can ``guess'' the correct neighbourhood of each vertex out of several options.
Additional constraints on the flow are to be satisfied to compensate for this guessing ability.
Using these techniques, it was possible to modify (in a rather intricate way) the welded tree graph achieving $O(n)$ quantum hitting time.
This is clearly optimal, as such is the distance between the opposite roots.

In this paper, we give a much simpler proof of the same optimal $O(n)$ upper bound on quantum hitting time on the welded tree graph.
Unlike Jeffery and Zur~\cite{jeffery:kDist}, and similarly to Li, Li, and Luo~\cite{li:weldedTreees}, we use the conventional version of discrete-time quantum walk on this graph.
Also, unlike Li, Li, and Luo, and similarly to Jeffery and Zur, we avoid any spectral analysis and use technique that is closely related to electric quantum walks.
Surprisingly, the only thing required to break the aforementioned quadratic barrier is to change the global phase of the walk!
The result is as follows, and is proven in \rf{sec:welded}.

\begin{thm}[Informal]
\label{thm:mainInformal}
$O(n)$ iterations of a standard discrete-time quantum walk on a welded tree graph of depth $n$ are sufficient to find the vertex of the opposite tree.
\end{thm}

The discrete-time quantum walk by Jeffery and Zur extensively uses the structure of the welded tree graph, and it is not apparent how to generalise it to other graphs.
Our algorithm can be applied to other graphs, which we demonstrate by analysing quantum hitting time on 1-dimensional hiearchical graphs.
\medskip

Unlike random walks, in quantum walks, there is a substantial difference between ability to detect presence of marked vertices, and actually finding one.
Szegedy's original paper~\cite{szegedy:walk} considered the problem of finding a marked vertex provided it is unique.
This case was later resolved by Krovi, Magniez, Ozols, and Roland~\cite{krovi:quantumSearch} using so-called interpolating quantum walk.
Ambainis, Gily\'{e}n, Jeffery, and Kokainis~\cite{ambainis:quadraticForFinding} showed that quantum search is as fast as detection for arbitrary number of marked vertices, provided the initial state is the steady state like in Szegedy's case.
Finally, Apers, Gily\'{e}n, and Jeffery~\cite{apers:unifiedQuantumWalk} obtained the same result for arbitrary initial distribution.

The proofs of these results are rather sophisticated.
Jeffery and Zur~\cite{jeffery:kDist} do not use any of these, but rely on a witty trick combining detection and the Bernstein-Vazirani algorithm.
Although this trick is not complicated and is very general, it does add an additional layer of sophistication to the algorithm.
We avoid this step, and solve the search problem directly.
In fact, we demonstrate an algorithm for finding unique marked vertex using electric quantum walk.
This algorithm is not as general the one by Apers, Gily\'{e}n, and Jeffery~\cite{apers:unifiedQuantumWalk}, but its analysis is much simpler.

We make use of a novel approach to quantum walks using transducers~\cite{belovs:taming}.
Although the same results can be obtained using more traditional phase estimation, we find the approach using transducers substantially cleaner.

The paper is organised as follows.
In \rf{sec:transducers}, we define transducers and their implementation by a quantum algorithm.
\rf{sec:electric} defines electric quantum walk, and sketches its implementation using diffuse-and-shift approach.
We modify the quantum walk slightly to capture the search problem.
In \rf{sec:weldedPrelim}, we define the welded tree problem.

In \rf{sec:walk}, we show how to solve the search problem using electric quantum walks provided the marked vertex is unique.
In \rf{sec:welded}, we prove \rf{thm:mainInformal}.
We note that it is not based on the walk in \rf{sec:walk}, the latter being given for comparison and to provide more intuition.
Finally, in \rf{sec:hierarchical}, we apply the same technique for the case of general 1-dimensional hierarchical graphs.

\section{Preliminaries}

\subsection{Transducers}
\label{sec:transducers}

The use of transducers~\cite{belovs:taming} is a novel approach to the design of quantum algorithms based on the ideas from~\cite{belovs:LasVegas}.
The following easy linear-algebraic result is the key to their definition.

\begin{thm}
\label{thm:introTransduce}
Let $S$ be a unitary that acts in a direct sum of two vectors spaces $\cH\oplus \cL$.
For every $\xi\in \cH$, there exist $\tau = \tau(S,\xi)\in \cH$ and $v = v(S,\xi)\in \cL$ such that
\begin{equation}
\label{eqn:1transduce}
S\colon \xi \oplus v \mapsto \tau \oplus v.
\end{equation}
Moreover, the vector $\tau$ is uniquely specified by $S$ and $\xi$, and the mapping $\xi \mapsto \tau(S,\xi)$ is unitary.
\end{thm}

In these settings, we call $S$ a transducer and say that $S$ \emph{transduces} $\xi$ into $\tau$, denoted $\xi\transduce{S} \tau$.
The motivation is that while $S$ does \emph{not} literally map $\xi$ into $\tau$, having $S$ is a legitimate way of implementing this transformation, as we show shortly.
The vector $v$ is called the \emph{catalyst} of this transduction, and $W(S,\xi) = \|v(S,\xi)\|^2$ is \emph{transduction complexity}.
The spaces $\cH$ and $\cL$ are called the \emph{public} and the \emph{private} spaces of the transducer $S$, respectively.
The following theorem states that we can efficiently perform transduction if we allow for a small error.

\begin{thm}
\label{thm:transducer}
Let vector spaces $\cH$, $\cL$, and a positive integer $K$ be fixed.
There exists a quantum algorithm that transforms $\xi$ into a vector $\tau'$ such that
\[
\|\tau' - \tau(S,\xi)\| \le 2 \sqrt{\frac{W(S,\xi)}{K}}
\]
for every transducer $S\colon \cH\oplus \cL\to\cH\oplus \cL$ and every initial state $\xi\in\cH$.
The algorithm uses $K$ controlled applications of $S$ and $O(K)$ other elementary operations.
\end{thm}

Therefore, if we allow a constant error, it suffices to execute the transducer $O(W)$ times, where $W$ is an upper bound on the transduction complexity $W(S,\xi)$.

\subsection{Electric quantum walks}
\label{sec:electric}
In this section, we define electric quantum walk on a graph, and sketch its implementation.
The analysis will be performed in \rf{sec:walk}.
We will be interested in the search problem, hence, our exposition will be slightly different from the one in~\cite{belovs:electicityQuantumWalks}.
For simplicity, we assume unique initial vertex.
The general case of an arbitrary initial probability distribution can be obtained in a way similar to~\cite{belovs:electicityQuantumWalks}.
The assumption of the unique \emph{marked} vertex is essential to the construction, though.

The walk takes place on a bipartite graph $G$ with parts $A$ and $B$, and the edge set $E$.
Each edge $e\in E$ is assigned a non-negative real weight $w_{e}$.
The walk starts in some vertex $s\in A$.
Exactly one vertex $t\in B$ is marked.
The goal is to find the marked vertex $t$.

For the quantum walk, we build the extended graph $G'$ as follows, see \rf{fig:walk}.
We add two new vertices $s'$ and $t'$ to the graph, as well as dangling edges $ss'$ and $t't$.%
\footnote{
The edge $ss'$ is adopted from~\cite{belovs:electicityQuantumWalks}.
The edge $t't$ is similar to the self-loop on a marked vertex in the interpolating quantum walk of Krovi \etal~\cite{krovi:quantumSearch} and the subsequent papers.
}
Note that $s'$ and $t'$ are neither in $A$, nor in $B$.
Let $E'$ be the set of the two newly added dangling edges.
Each edge in $E'$ has weight 1.
%The edge $ss'$ is as in~\cite{belovs:electicityQuantumWalks}.

\myfigure{\label{fig:walk}}
{
An example of the extension of a graph for a quantum walk.
The original graph contains two parts $A$ and $B$ of 4 and 3 vertices, respectively.
They are marked yellow and green, respectively.
The initial vertex $s\in A$ is attached a dangling edge $ss'$, and the marked vertex $t\in B$ is attached a dangling edge $t't$.
The vertices $s'$ and $t'$ are neither in $A$, nor in $B$.
All the edges (including the new ones) are oriented from $A$ and/or towards $B$.
The weights of the original edges are given by $w_e$, the weights of the new edges are 1.
}
{
\negbigskip
\[
\begin{tikzpicture}[
    minimum size=18pt, 
    inner sep=1pt,
    inA/.style={circle, draw, fill=yellow!30},
    inB/.style={circle, draw, fill=green!30},    
]
    \node[black!80] at (0,-0.8) {\Large $A$};
    \node[inA] (A4) at (0,0) {};
    \node[inA] (A3) at (0,1) {};
    \node[inA] (A2) at (0,2) {};
    \node[inA] (A1) at (0,3) {$s$};
    \node[black!80] at (2.5,-0.8) {\Large $B$};
    \node[inB] (B3) at (2.5,0.5) {};
    \node[inB] (B2) at (2.5,1.5) {$t$};
    \node[inB] (B1) at (2.5,2.5) {};
    \graph{
        (A1)->{(B1),(B3)};
        (A2)->{(B2),(B3)};
        (A3)->{(B1),(B2),(B3)};
        (A4)->{(B1),(B2)};
    };
    \draw (A1) to node[above]{$w_e$} (B1);
    \node[circle, draw, fill=\markedcolor!50] (M) at (4,1.5) {$t'$};
    \node[circle, draw, fill=\inputcolor!50] (S1) at (-1.5,3) {$s'$};
    \draw[<-,\inputcolor] (S1) to node[above] {$1$} (A1);
    \draw[<-,\markedcolor] (B2) to node[above] {$1$} (M);
\end{tikzpicture}
\]
\negbigskip
}

The space of the quantum walk is $\bC^{E\cup E'}$, where we treat $\bC^{E'}$ and $\bC^E$ as the public and the private space, respectively, in the sense of \rf{sec:transducers}.
For brevity, we will denote $\ket |s> = \ket |ss'>$ and $\ket |t> = \ket |t't>$.
We assume each edge $\ket |e>$ is given some orientation.
Precise choice of orientation is not important, as its change merely results in change of the sign: $\ket |uv> = -\ket |vu>$.
For concreteness, however, we will assume all the edges are oriented towards the vertices in $B$ and away from the vertices in $A$.

For each $u\in A\cup B$, define the following \emph{star state}:
\begin{equation}
\label{eqn:psi_u}
\psi_u = 
\sum_{e\sim u} \sqrt{w_{e}} \ket |e>,
\end{equation}
where the sum is over all the edges incident to $u$ (including the new edges from $E'$).
We note that this simple form of the state $\psi_u$ is due to our choice of the orientation: either all the edges are oriented towards $u$, or from $u$.
Should the edges vary in direction, this would have been represented by different signs in~\rf{eqn:psi_u}.

\begin{defn}
\emph{One step (or iteration) of the quantum walk} on the above extended graph $G'$ is defined as $U = U(G') = R_B(G')R_A(G')$, where $R_A = R_A(G')$ (resp., $R_B = R_B(G')$) denotes the reflection of all $\psi_u$ where $u$ ranges over $A$ (resp., $B$).
In other words, $R_A$ is defined as negation on the span of $\{\psi_u\}_{u\in A}$ and identity on its orthogonal complement, and $R_B$ is defined similarly for $\{\psi_u\}_{u\in B}$.
\end{defn}

Sometimes we talk about quantum walk on $G$ when we actually mean quantum walk on $G'$.

Let us make some notes on the implementation of $U$.
In general, we need not assume the algorithm knows the whole graph.
The following three primitives are essential, however:
\begin{itemize}\itemsep=0pt
\item The algorithm knows the initial vertex $s$.
\item For any $u\in A\cup B$, the algorithm can test whether $u$ is marked.
\item For any $u\in A\cup B$, the algorithm can learn the neighbourhood of $u$.
More specifically, it is possible to prepare the state $\psi_u$ from~\rf{eqn:psi_u} in order to implement $R_A$ and $R_B$.
\end{itemize}

It may seem one has to know whether a vertex $u$ is in $A$ or $B$ in order to understand which $\psi_u$ to reflect.
But this can be circumvented using the following standard construction.
Let $K_2$ be the complete graph on two vertices $0$ and $1$.
Consider the tensor product $G\times K_2$ of the graphs.
Its vertex set is $V\times\{0,1\}$ with $V=A\cup B$ being the vertex set of $G$.
There is an edge between $(u,a)$ and $(v,b)$ in $G\times K_2$ iff $uv$ is an edge of $G$, and $a\ne b$.
The weight of this edge is equal to the weight of $uv$ in $G$.

The graph $G\times K_2$ is bipartite with parts $V\times\{0\}$ and $V\times \{1\}$ even if $G$ is not bipartite.
If $G$ is bipartite, $G\times K_2$ breaks down into two disconnected components, each isomorphic to $G$: the first one with the parts $A\times\{0\}$ and $B\times\{1\}$, and the second one with the parts $A\times\{1\}$ and $B\times\{0\}$.
A (quantum) walk on the graph $G_2$ with the initial vertex $(s,0)$ only takes place in the first component, while the second one is inaccessible and can be ignored.
Therefore, in this case, quantum walks on $G$ and $G\times K_2$ are identical.

Exact details on how the edges are represented in memory and how the states $\psi_u$ are prepared depend on particular application.
The following diffuse-and-shift approach is quite common, though.
The space of the walk is embedded into the space spanned by $\ket|u>\ket|v>$ with $u,v$ in the vertex set $V$ of $G$.
The state $\ket|u>\ket |v>$ represents the edge between $(u,0)$ and $(v,1)$ in $G\times K_2$.
The quantum walk alternates between the diffuse operation $D$, and the shift operation $S$.
The diffuse operation performs the reflection of the states
\[
\psi_{(u,0)} = \ket |u>\otimes \sum_{v} \sqrt{w_{uv}} \ket |v>
\]
over all $u\in V$, where the sum is over all the vertices $v$ incident to $u$.
The shift operation swaps the two registers.
It is easy to see that $D=R_{V\times\{0\}}$ and $SDS = R_{V\times\{1\}}$, hence, $U = SDSD$ equals one step of the quantum walk on $G\times K_2$.
%In the case of the welded tree graph, the diffuse operation $D$ can be implemented using an oracle that returns the list of neighbours of a vertex, see~\cite[Appendix A]{li:weldedTreees}.

The above discussion neglected the new dangling edges $ss'$ and $t't$.
They can be represented as $\ket |s>\ket |\perp>$ and $\ket|\perp>\ket |t>$, where $\perp$ is some specific element outside of $V$.
The diffuse operation should be modified accordingly.
This can be done as the algorithm knows the vertex $s$, and can test whether a particular vertex is marked.

\def\weldedtree{
    \begin{scope}
    [
        inA/.style={fill=yellow!30},
        inB/.style={fill=green!30},
        big/.style={inner sep=\bignodesize pt},
        every node/.style={circle, draw, inner sep=3pt},
        level distance=1cm
    ]
        \node[big, fill=\inputcolor!50] (S') {} 
        child[<-, grow=right] {
            node[big, inA] (S) {}
            child [->, sibling distance=26mm] foreach \x in {0,1}{ 
                node[inB] (S\x){}
                child [<-, sibling distance=13mm] foreach \y in {0,1}{
                    node[inA] (S\x\y) {}
                    child [->, sibling distance=7mm] foreach \z in {0,1}{
                        node[inB] (S\x\y\z) {}
                    }
                }
            }
        };
        \node[big, fill=\markedcolor!50] at (9,0) (T') {} 
        child[->, grow=left] {
            node[big, inB] (T) {}
            child [<-, sibling distance=26mm] foreach \x in {0,1}{ 
                node[inA] (T\x){}
                child [->, sibling distance=13mm] foreach \y in {0,1}{
                    node[inB] (T\x\y) {}
                    child [<-, sibling distance=7mm] foreach \z in {0,1}{
                        node[inA] (T\x\y\z) {}
                    }
                }
            }
        };
    \end{scope}
    \graph { 
    (S111)<-(T000)->
    (S110)<-(T010)->
    (S100)<-(T100)->
    (S010)<-(T110)->
    (S000)<-(T111)->
    (S001)<-(T101)->
    (S011)<-(T011)->
    (S101)<-(T001)->
    (S111);};
    \node at (S') {$s'$};
    \node at (S) {$s$};
    \node at (T) {$t$};
    \node at (T') {$t'$};
}
\def\weldedtreewithlabels S' #1 S #2 S1 #3 S2 #4 S3 #5 T3 #6 T2 #7 T1 #8 T #9 T'{
    \weldedtree
    \draw [every node/.style={auto, blue,font={\scriptsize}}] 
    (S')--node{$#1$}(S)--node{$#2$}(S1)--node{$#3$}(S11)--node{$#4$}(S111)--node{$#5$}(T000)--node{$#6$}(T00)--node{$#7$}(T0)--node{$#8$}(T)--node{$#9$}(T');
}

\subsection{Welded Tree Problem}
\label{sec:weldedPrelim}

A welded tree graph $G_n$ of depth $n$ consists of two complete binary trees of depth $n$ with a cycle between the leaves of the trees, see \rf{fig:weldedtree}.
The cycle uses each of the leaves exactly once, and only uses edges between leaves from two different trees.
We call one tree left, and the other one right.
For each tree, we divide the edges into layers, where layer $1$ is incident to the root, layer $n$ is incident to the leaves, and, in general, the edges in layer $i$ connect vertices at distance $i-1$ to vertices at distance $i$ from the root.
The edges connecting the leaves of both trees comprise the middle layer.
Each edge of the graph has weight 1.

The root of the left tree is the initial vertex $s$, and the root of the right tree is the marked vertex $t$.
This is a bipartite graph with parts we denote $A$ and $B$.
The initial vertex $s$ is in $A$, and the marked vertex $t$ is in $B$, as required in \rf{sec:electric}.
We modify the graph as described in the same section by adding two dangling edges $ss'$ and $t't$, both of weight 1.

\myfigure{\label{fig:weldedtree}}
{
An example of a welded tree graph with depth $n=3$ extended with two dangling edges.
The colour coding is similar to \rf{fig:walk}: the vertices in $A$ are yellow, and the ones in $B$ are green.
The orientation of the edges is also the same.
The layer $i$ of the left tree is marked by $L_i$, that of the right tree by $R_i$, and the middle layer is denoted by $M$.
}
{
\[
\def\bignodesize{4.5}
\begin{tikzpicture}
\weldedtreewithlabels S' {}  S L_1 S1 L_2 S2 L_3 S3 M T3 R_3 T2 R_2 T1 R_1 T {} T'
\end{tikzpicture}
\]
}

In the welded tree problem one has to find the node $t$ given the node $s$.
The graph is accessible via an oracle that, given a vertex $u$ in the graph, outputs a list of its three neighbours (or two neighbours in the case of $s$ and $t$).
With this oracle, it is easy to implement quantum walk on the extended graph $G_n'$ using the diffuse-and-shift technique from the end of \rf{sec:electric}.
For instance, it is possible to use the same algorithm as in~\cite[Appendix A]{li:weldedTreees} with minor modification to include the new dangling edges $ss'$ and $t't$.
Even more, their addition makes the graph more uniform, as now all the original vertices of the graph have degree 3.

\section{Simple Search Using Electric Quantum Walks}
\label{sec:walk}

In this section, we give a simple extension of the electric quantum walk from~\cite{belovs:electicityQuantumWalks} that finds a marked vertex provided it is unique.
For simplicity, we assume that the initial vertex of the quantum walk is also unique.
These conditions are satisfied in the welded tree problem.

Let
\[
W = \sum_{e\in E} w_e
\]
be the total weight of the graph, and let $R_{s,t}$ be the minimum of
\begin{equation}
\label{eqn:resistance}
\sum_{e\in E} \frac{p_e^2}{w_e},
\end{equation}
over all unit flows $p=(p_e)$ from $s$ to $t$ in the graph $G$.
The minimum is attained by the electrical flow, and $R_{s,t}$ is the corresponding effective resistance.

\begin{thm}
\label{thm:walkTechnical}
One step $U = R_BR_A$ of the quantum walk transduces $\ket |s>$ into $\ket |t>$ with transduction complexity at most $\frac12(W + R_{s, t})$.
\end{thm}

Assuming we know an upper bound $W$ on the total weight, and $R$ on the effective resistance, we can get an algorithm that finds $t$ in $O(\sqrt{WR})$ iterations of the quantum walk using the following standard re-weighting trick.
Let $\alpha = \sqrt{R/W}$.
We convert $G$ into the graph $\alpha G$ by multiplying the weight of each edge $e\in E$ by $\alpha$.
The new graph $\alpha G$ has total weight upper bounded by $\alpha W$ and the effective resistance by $R/\alpha$, which are both equal to $\sqrt{RW}$.
Now we can apply Theorems~\ref{thm:walkTechnical} and~\ref{thm:transducer}.

Note that application of \rf{thm:transducer} introduces error, while transduction in \rf{thm:walkTechnical} is exact.
The latter can be useful if the transducer $U$ is composed with other transducers, see~\cite{belovs:taming}.
The same applies to other transducers in this paper.

\begin{proof}[Proof of \rf{thm:walkTechnical}]
We will show that $\ket |s> - \ket |t> \transduce{U} -\ket |s> + \ket |t>$ and $\ket |s> + \ket |t> \transduce{U} \ket |s> + \ket |t>$, and obtain the final transduction by taking the linear combination of the two.
Both of these transductions closely follow the corresponding constructions for detection from~\cite{belovs:taming}.

For the first transduction, the catalyst is the following sum of the edges:
\[
\ket |v_0> = \sum_{e\in E} \sqrt{w_e} \ket |e>.
\]
Note that
\[
\ket |s> + \ket |v_0> = \sum_{u\in A} \psi_u
\qqand
\ket |v_0> + \ket |t> = \sum_{u\in B} \psi_u,
\]
hence, by the definition of $R_A$ and $R_B$, we get
\begin{equation}
\label{eqn:walkTech1}
\ket |s> + \ket |v_0> - \ket |t> 
\maps{R_A} 
-\ket |s> - \ket |v_0> - \ket |t> 
\maps{R_B}
-\ket |s> + \ket |v_0> + \ket |t>.
\end{equation}

For the second transduction, the catalyst is given by the flow:
\[
\ket |v_1> = \sum_{e\in E} \frac {p_e}{\sqrt {w_e}} \ket |e>.
\]
Here $p = (p_e)$ is the unit flow from $s$ to $t$ in $G$ attaining the minimum in~\rf{eqn:resistance}.
This flow can be extended to the flow from $s'$ to $t'$ in the extended graph $G'$ by letting $p_{ss'} = p_{t't} = -1$ (as the flow goes in the direction opposite to the orientation of the edges).
The flow is preserved in each $u\in A\cup B$, therefore
\[
-\ket |s> + \ket |v_1> - \ket |t>  \perp \spn_{u\in A\cup B} \psi_u,
\]
and
\begin{equation}
\label{eqn:walkTech2}
-\ket |s> + \ket |v_1> - \ket |t> 
\maps{R_A}
-\ket |s> + \ket |v_1> - \ket |t>
\maps{R_B}
-\ket |s> + \ket |v_1> - \ket |t>.
\end{equation}

By subtracting~\rf{eqn:walkTech2} from~\rf{eqn:walkTech1}, we get
\[
\ket |s> + \frac{v_0 - v_1}{2} \maps{U} \ket |t> + \frac{v_0 - v_1}{2}.
\]
Thus, $U$ transduces $\ket |s>$ into $\ket |t>$ with transduction complexity
\[
\frac{\|v_0 - v_1\|^2}{4} 
\le 
\frac12\sA[\|v_0\|^2 + \|v_1\|^2]
=
\frac12\sA[W+R_{s,t}].\qedhere
\]
\end{proof}

\section{Welded Trees}
\label{sec:welded}

First, we will show that the standard electric quantum walk of \rf{thm:walkTechnical} requires exponential (in $n$) number of iterations to find $t$.
After that, we introduce a simple modification that reduces the required number of iterations all the way down to $O(n)$.

\myfigure{\label{fig:welded1Witnesses}}
{
The vectors used in the proof of \rf{thm:walkTechnical} for the case of the graph in \rf{fig:weldedtree}.
All the edges in the same layer have the same amplitude that is given above the highest edge of the layer.
The figure on the left is the sum of the edges (except for the negated $\ket |t>$).
The one on the right represents the flow from $s'$ to $t'$ of value 1.
In a general welded tree graph, the amplitude is $\pm1$ divided by the number of edges in the corresponding layer.
}
{
\negbigskip
\[
\hspace{-.45cm}
\def\bignodesize{3}
\begin{array}{ccc}
\ket |s> + \ket |v_0> - \ket |t>  & -\ket |s> + \ket |v_1> - \ket |t> \\
\begin{tikzpicture}[scale=0.8]
\weldedtreewithlabels S' 1 S 1 S1 1 S2 1 S3 1 T3 1 T2 1 T1 1 T -1 T'
\end{tikzpicture}
&
\begin{tikzpicture}[scale=0.8]
\weldedtreewithlabels S' -1 S \frac 12 S1 -\frac 14 S2 \frac 18 S3 -\frac1{16} T3
 \frac 18 T2 -\frac 14 T1 \frac 12 T -1 T'
\end{tikzpicture}
\end{array}
\]
}

The states $\ket |s> + \ket |v_0> - \ket |t>$ from~\rf{eqn:walkTech1} and $-\ket |s> + \ket |v_1> - \ket |t>$ from \rf{eqn:walkTech2} for this particular case are depicted in \rf{fig:welded1Witnesses}.
It is not hard to see that
\begin{equation}
\label{eqn:v0 and v1}
\|v_0\|^2 = 2 \sum_{i=1}^n 2^{i}\cdot 1 + 2^{n+1}\cdot 1 = \Theta(2^n)
\quad\text{and}\quad
\|v_1\|^2 = 2 \sum_{i=1}^n 2^{i}\frac{1}{2^{2i}} + 2^{n+1}\frac{1}{2^{2(n+1)}} = \Theta(1).
\end{equation}
By \rf{thm:walkTechnical}, this results in an exponential number of iterations.

The problem here is that the catalysts $v_0$ and $v_1$ are not balanced.  The coefficients in the first sum in~\rf{eqn:v0 and v1} decrease too slowly (rather, do not decrease at all), while the ones in the second sum decrease even too fast.
We solve this problem by considering a mixture of these two catalysts.
For simplicity, we assume $n$ is odd.  
The case of even $n$ is similar, but slightly more complicated,
and is captured by a more general \rf{thm:hierarchical}.

\myfigure{\label{fig:welded2Witnesses}}
{
The vectors used in the proof of \rf{thm:welded}.
All the edges in the same layer have the same amplitude that is given above the highest edge of the layer.
Both vectors behave like the sum of the edges for vertices from one part of the graph, and as a flow for the other part.
}
{
\negbigskip
\[
\hspace{-.45cm}
\def\bignodesize{3}
\begin{array}{ccc}
\ket |s> + \ket |v_2> - \ket |t>  & \ket |s> + \ket |v_3> + \ket |t> \\
\begin{tikzpicture}[scale=0.8]
\weldedtreewithlabels S' 1 S  1 S1 -\frac 12 S2 -\frac 12 S3 \frac14 T3
 \frac 14 T2 -\frac 12 T1 -\frac 12 T -1 T'
\end{tikzpicture}
&
\begin{tikzpicture}[scale=0.8]
\weldedtreewithlabels S' 1 S  -\frac 12 S1 -\frac 12 S2 \frac 14 S3 \frac14 T3
 -\frac 12 T2 -\frac 12 T1 1 T 1 T'
\end{tikzpicture}
\end{array}
\]
}

\begin{thm}
\label{thm:welded}
Let $U = R_BR_A$ be one step of the quantum walk on the welded tree graph of odd depth $n$.
Then  $-U$ transduces $\ket |s>$ into $\ket |t>$ with transduction complexity at most $3(n+1)$.
\end{thm}

By \rf{thm:transducer}, it takes $O(n)$ iterations of the quantum walk to find $t$.
Surprisingly, the only difference between operators used in Theorems~\ref{thm:walkTechnical} and~\ref{thm:welded} is the global phase, but it results in performance estimates that are exponentially far apart!

\begin{proof}[Proof of \rf{thm:welded}]
Consider the following vector $v_2\in \bC^E$, see also \rf{fig:welded2Witnesses}:
\begin{equation}
\label{eqn:v2}
v_2\elem[e] = \begin{cases}
(-1/2)^i, & \text{if $e$ belongs to level $2i$ or $2i+1$ of the left tree;}\\
(-1/2)^i, & \text{if $e$ belongs to level $2i$ or $2i-1$ of the right tree.}
\end{cases}
\end{equation}
The middle layer between the trees can be considered as the $(n+1)$-st level of both trees, and both cases of~\rf{eqn:v2} apply there (we are using that $n$ is odd here).
This vector behaves as the sum of the edges for vertices in $A$, and as a flow for vertices in $B$.
In other words,
\[
\ket |s> + \ket |v_2> \in \spn_{u\in A} \psi_u
\qqand
\ket |v_2> + \ket |t> \perp \spn_{u\in B} \psi_u.
\]
This gives us
\begin{equation}
\label{eqn:weldedTech1}
\ket |s> + \ket |v_2> - \ket |t> 
\maps{R_A} 
-\ket |s> - \ket |v_2> - \ket |t> 
\maps{R_B}
-\ket |s> - \ket |v_2> - \ket |t>.
\end{equation}
Similarly, we define $v_3\in \bC^E$, which is a mirror image of $v_2$:
\begin{equation*}
%\label{eqn:v3}
v_3\elem[e] = \begin{cases}
(-1/2)^i, & \text{if $e$ belongs to level $2i$ or $2i-1$ of the left tree;}\\
(-1/2)^i, & \text{if $e$ belongs to level $2i$ or $2i+1$ of the right tree.}
\end{cases}
\end{equation*}
Again, the middle layer is covered by both cases.
However, this time we have
\[
\ket |s> + \ket |v_2> \perp \spn_{u\in A} \psi_u,
\qqand
\ket |v_2> + \ket |t> \in \spn_{u\in B} \psi_u,
\]
and
\begin{equation}
\label{eqn:weldedTech2}
\ket |s> + \ket |v_3> + \ket |t> 
\maps{R_A} 
\ket |s> + \ket |v_3> + \ket |t> 
\maps{R_B}
\ket |s> - \ket |v_3> - \ket |t>.
\end{equation}
Combining~\rf{eqn:weldedTech1} and~\rf{eqn:weldedTech2}, we get for $U = R_BR_A$:
\begin{equation}
\label{eqn:weldedTransduction}
\ket |s> + \frac{v_2 + v_3}{2} \maps{-U} \ket |t> + \frac{v_2 + v_3}{2},
\end{equation}
which gives the required transduction.
We can estimate the norms of $v_2$ and $v_3$ as 
two times the heaviest tree including the middle layer:
\begin{equation}
\label{eqn:v23norms}
\|v_2\|^2 = \|v_3\|^2 \le 2 \sum_{i=1}^{(n+1)/2} \sB[ 2^{2i-1}\cdot\frac{1}{2^{2(i-1)}} + 2^{2i}\cdot \frac 1{2^{2i}} ] = 3(n+1).
\end{equation}
From this, the bound on transduction complexity follows, as
\[
\frac{\|v_2 + v_3\|^2}{4} 
\le 
\frac12\sA[\|v_2\|^2 + \|v_3\|^2]
\le 3(n+1).\qedhere
\]
\end{proof}

\section{Hierarchical 1-Dimensional Graphs}
\label{sec:hierarchical}
In this section, we apply the technique from \rf{thm:welded} for 1-dimensional hierarchical graphs.
They are defined in~\cite{balasubramanian:hierarchicalGraphs}, and form a generalisation of all the examples mentioned in the introduction: the welded tree graph, the glued tree graph, and the hypercube.

\begin{defn}
The vertex set of a \emph{1-dimensional hierarchical graph} $G$ is a disjoint union $S_0\sqcup S_1\sqcup\cdots\sqcup S_n$ with $|S_0| = |S_n| = 1$.
The edge set is a disjoint union $L_1\sqcup L_2\sqcup \cdots \sqcup L_n$.
The edges in $L_i$ form a bipartite graph with parts $S_{i-1}$ and $S_i$, where each vertex in $S_{i-1}$ has degree $d_{i-1}^{+}$, and each vertex in $S_i$ has degree $d_i^{-}$.

In the corresponding \emph{search problem} one has to find the sole vertex $t$ in $S_n$ given the sole vertex $s$ in $S_0$.
It is assumed that there is an oracle that can tell whether a vertex $v$ equals $t$.
\end{defn}

The graph can only exist if $|S_{i-1}| d_{i-1}^+ = |S_i|d_i^- = |L_i|$.
We will assume that $n$ is odd.%
\footnote{
The case of even $n$ can be reduced to the odd case by adding one more layer of edges $E_{n+1}$ and vertices $S_{n+1}$ with $|E_{n+1}| = |S_{n+1}| = 1$.}
The graph $G$ is bipartite with parts $A = S_0 \cup S_2 \cup \cdots \cup S_{n-1}$ and $B= S_1\cup S_3\cup \cdots \cup S_n$.
We will use discrete-time quantum walk from \rf{sec:electric}.
In particular, we extend the graph to $G'$ by adding two dangling edges $ss'$ and $t't$.
We assume the weights of all the original edges in $G$ are 1.
The weights of the edges $ss'$ and $t't$ are $W_0$ and $W_{n+1}$, respectively.
We do not assume they are equal to 1 here, contrary to \rf{sec:electric}.
We also denote $W_i = |L_i|$ for all $i\in\{1,2,\dots,n\}$.

%In this section, 
We assume we can implement the transformation $U = R_BR_A$ efficiently, and we will be interested in the number of applications of this unitary required to solve the search problem.
Let
\[
C_k = \prod_{i=1}^{k} \s[\frac{W_{i-1}}{W_{i}}]^{(-1)^{i}} = 
\frac{W_1}{W_0}\cdot \frac{W_1}{W_2}\cdot \frac{W_3}{W_2} \cdot\frac{W_3}{W_4} \cdot\cdots\cdot \s[\frac{W_{k-1}}{W_{k}}]^{(-1)^{k}}.
\]
Assuming the graph $G$ is fixed, for each choice of the weight $W_0$ of $ss'$, there is unique choice of the weight $W_{n+1}$ of $t't$ such that $C_{n+1}=1$.
The main result of this section is the following theorem, which is comparable to a similar result in~\cite{balasubramanian:hierarchicalGraphs}.

\begin{thm}
\label{thm:hierarchical}
Let $G'$ be an extended 1-dimensional hierarchical graph as above such that $C_{n+1}=1$.
Denote by $U = R_BR_A$ one step of the quantum walk on this graph.
Then  $-U$ transduces $\ket |s>$ into $(-1)^{(n+1)/2}\ket |t>$ with transduction complexity not exceeding
\[
\frac12\sC[\sum_{k=1}^n {\sB[C_k + \frac{1}{C_k}]}].
\]
\end{thm}

\def\linegraph#1#2#3#4#5#6#7#8#9{
    \begin{scope}
    [
        vertex/.style={circle, draw, inner sep=2pt},
        inA/.style={vertex, fill=yellow!30},
        inB/.style={vertex, fill=green!30},
    ]
    \matrix [column sep=9mm, ampersand replacement=\&]
    {
        \node[vertex, fill=\inputcolor!50] (S') {$s'$}; \&
        \node[inA] (S) {$s_0$}; \&
        \node[inB] (S1) {$s_1$}; \&
        \node[inA] (S2) {$s_2$}; \&
        \node[inB] (S3) {$s_3$}; \&
        \node[inA] (S4) {$s_4$}; \&
        \node[inB] (S5) {$s_5$}; \&
        \node[inA] (S6) {$s_6$}; \&
        \node[inB] (T) {$s_7$}; \&
        \node[vertex,fill=\markedcolor!50] (T') {$t'$}; \\
    };
    \end{scope}
%    \draw (S') <- (S) -> (S1) <- (S2) -> (S3) <- (S4) -> (S5) <- (S6) -> (T) <- (T');
    \draw[<-] (S') --node[above, blue]{$#1$}node[below]{1} (S);
    \draw[->] (S) --node[above, blue]{$#2$}node[below]{$2$} (S1);
    \draw[<-] (S1) --node[above, blue]{$#3$}node[below]{$4$} (S2);
    \draw[->] (S2) --node[above, blue]{$#4$}node[below]{$8$} (S3);
    \draw[<-] (S3) --node[above, blue]{$#5$}node[below]{$16$} (S4);
    \draw[->] (S4) --node[above, blue]{$#6$}node[below]{$8$} (S5);
    \draw[<-] (S5) --node[above, blue]{$#7$}node[below]{$4$} (S6);
    \draw[->] (S6) --node[above, blue]{$#8$}node[below]{$2$} (T);
    \draw[<-] (T) --node[above, blue]{$#9$}node[below]{1} (T');
}

\myfigure{\label{fig:flatted}}
{
The flattened version of the graph from \rf{fig:weldedtree}.
The vertices in $\bar A$ are yellow, and the ones in $\bar B$ are green.
Below an edge, its weight is given. Here, both $W_0$ and $W_{n+1}$ are equal to 1.
\\
(a) The correspondence between the edges and the corresponding vectors.
\\
(b) The vector $\ket |s> + \ket |v_2> - (-1)^{(n+1)/2}\ket |t>$.
(c) The vector $\ket |s> + \ket |v_3> + (-1)^{(n+1)/2}\ket |t>$.
\\
In both (b) and (c), the numbers above the edges are amplitudes.
These vectors are equivalent to the ones in \rf{fig:welded2Witnesses}.
}
{
\negbigskip
\[
\def\bignodesize{4.5}
\begin{tikzpicture}
\linegraph{\ket|L_0>}{\ket|L_1>}{\ket|L_2>}{\ket|L_3>}{\ket|L_4>}{\ket|L_5>}{\ket|L_6>}{\ket|L_7>}{\ket|L_8>}
\draw (S') +(-1,0) node {(a)};
\end{tikzpicture}
\]
\[
\def\bignodesize{4.5}
\begin{tikzpicture}
\linegraph{1}{\sqrt{2}}{-1}{-\sqrt{2}}{1}{\sqrt{\frac 12}}{-1}{-\sqrt{\frac12}}{-1}
\draw (S') +(-1,0) node {(b)};
\end{tikzpicture}
\]
\[
\def\bignodesize{4.5}
\begin{tikzpicture}
\linegraph{1}{-\sqrt{\frac12}}{-1}{\sqrt{\frac12}}{1}{-\sqrt{2}}{-1}{\sqrt{2}}{1}
\draw (S') +(-1,0) node {(c)};
\end{tikzpicture}
\]
}

Before we prove the theorem, let us explain the standard technique of reducing a quantum walk on $G_n$ to a quantum walk on the line, utilised in pretty much all previous papers~\cite{kempe:walkHitFaster, childs:gluedTrees, childs:walkExponentialSeparation, li:weldedTreees, balasubramanian:hierarchicalGraphs}.
Let
\[
\ket |L_i> = \frac1{\sqrt{|L_i|}} \sum_{e\in L_i} \ket |e>.
\]
We define the \emph{flattened graph} $\bar G$ as follows.
It has vertices $s_0,s_1,\dots, s_n$.
The edge between $s_{i-1}$ and $s_i$ is given by $\ket |L_i>$, and it has weight $W_i = |L_i|$.
This graph is bipartite with parts 
$\bar A = \{s_0, s_2, \dots s_{n-1}\}$
and
$\bar B = \{s_1, s_3, \dots s_n\}$.
Again, we extend the graph to $\bar G'$ by adding edges $\ket |L_0> = \ket |s> = \ket |ss'>$ and $\ket |L_{n+1}> = \ket |t> = \ket|t't>$ incident to $s_0$ and $s_n$, respectively.
They have the same weight as in $G'$: $W_0$ and $W_{n+1}$, respectively.
See \rf{fig:flatted}(a) for an example.

\begin{prp}
\label{prp:flat}
Both operations $R_A(G')$ and $R_B(G')$ of the quantum walk step on the graph $G'$, when restricted to the space spanned by $\ket|L_0>,\ket|L_1>,\cdots, \ket|L_{n+1}>$, are identical to the operations $R_{\bar A}(\bar G')$ and $R_{\bar B}(\bar G')$ of the quantum walk on the graph $\bar G'$.
\end{prp}

\begin{proof}
For $i\in\{0,\dots,n\}$, we have
\begin{equation}
\label{eqn:line1}
\sum_{u \in S_i} \ket|\psi_u> 
%= 
= \sqrt{W_i}\; \ket |L_i> + \sqrt{W_{i+1}}\; \ket |L_{i+1}>
= \ket|\psi_{s_i}>,
\end{equation}
where the star states of the left are from $G'$ and on the right from $\bar G'$
Indeed, both sides are equal to
\[
\sqrt{W_0} \ket |ss'> + \sum_{e\in L_1} \ket |e>,
\qquad
\sum_{e\in L_{i-1}\cup L_i} \ket |e>,
\qquad\text{or}\qquad
\sum_{e\in L_n} \ket |e> + \sqrt{W_{n+1}} \ket |t't>
\]
for $i=0$, $0<i<n$, and $i=n$, respectively.

Also,
$
\sqrt{W_{i+1}}\; \ket |L_i> - \sqrt{W_{i}}\; \ket |L_{i+1}>
$
is orthogonal to~\rf{eqn:line1}.
By symmetry, it has the same inner product with all $\psi_u$ for $u\in S_i$, hence, it is orthogonal to all of them.
Therefore, in the space spanned by $\ket|L_{i}>$ and $\ket|L_{i+1}>$, the reflection of all $\psi_u$ with $u\in S_i$ is identical to the reflection of~$\psi_{s_i}$.
The proposition follows from the definitions of $R_A$ and $R_B$.
\end{proof}

\begin{proof}[Proof of \rf{thm:hierarchical}]
The proof is similar to that of \rf{thm:welded}.  
First, we can use \rf{prp:flat}, and analyse the quantum walk on the graph $\bar G'$ instead.
Define
\[
\ket|v_2> = \sum_{k=1}^{n} (-1)^{\floor[k/2]} \sqrt{C_k} \ket |L_k>,
\]
where $\floor[x]$ is the floor operation.
We have (where we assume $C_0=1$):
\[
\ket |s> + \ket |v_2> = \sum_{j=0}^{(n-1)/2} (-1)^j \sqrt{C_{2j}} 
\s[\ket |L_{2j}> + \sqrt{\frac{W_{2j+1}}{W_{2j}}}\ket|L_{2j+1}> ] \in \spn_{v \in \bar A} \psi_{v}.
\]
On the other hand,
\[
\ket |v_2> + (-1)^{(n+1)/2}\ket |t> = \sum_{j=1}^{(n+1)/2} (-1)^{j-1} \sqrt{C_{2j-1}} 
\s[\ket |L_{2j-1}> - \sqrt{\frac{W_{2j-1}}{W_{2j}}}\ket|L_{2j}> ] \perp \spn_{v \in \bar B} \psi_{v}.
\]
As in~\rf{eqn:weldedTech1}, this gives us
\begin{equation}
\label{eqn:hierTech1}
\ket |s> + \ket |v_2> - (-1)^{(n+1)/2} \ket |t> 
\maps{R_{\bar A}} 
-\ket |s> - \ket |v_2> - (-1)^{(n+1)/2} \ket |t> 
\maps{R_{\bar B}}
-\ket |s> - \ket |v_2> - (-1)^{(n+1)/2} \ket |t>.
\end{equation}
Define also
\[
\ket|v_3> = \sum_{k=1}^{n} (-1)^{\ceil[k/2]} \sqrt{\frac1{C_k}} \ket |L_k>,
\]
where $\ceil[x]$ is the ceiling operation.
Again, we have:
\[
\ket |s> + \ket |v_3> = \sum_{j=0}^{(n-1)/2} (-1)^j \sqrt{\frac1{C_{2j}}} 
\s[\ket |L_{2j}> - \sqrt{\frac{W_{2j}}{W_{2j+1}}}\ket|L_{2j+1}> ] \perp \spn_{v \in \bar A} \psi_{v},
\]
and
\[
\ket |v_3> + (-1)^{(n+1)/2}\ket |t> = \sum_{j=1}^{(n+1)/2} (-1)^{j} \sqrt{\frac1{C_{2j-1}}} 
\s[\ket |L_{2j-1}> + \sqrt{\frac{W_{2j}}{W_{2j-1}}}\ket|L_{2j}> ] \in \spn_{v \in \bar B} \psi_{v}.
\]
As in~\rf{eqn:weldedTech2}, this gives us
\begin{equation}
\label{eqn:hierTech2}
\ket |s> + \ket |v_3> + (-1)^{(n+1)/2}\ket |t> 
\maps{R_A} 
\ket |s> + \ket |v_3> + (-1)^{(n+1)/2}\ket |t> 
\maps{R_B}
\ket |s> - \ket |v_3> - (-1)^{(n+1)/2}\ket |t>.
\end{equation}
Combining~\rf{eqn:hierTech1} and~\rf{eqn:hierTech2}, and using \rf{prp:flat}, we get for both $U = R_B(\bar G')R_A(\bar G')$ and $U = R_B(G')R_A(G')$:
\begin{equation}
\ket |s> + \frac{v_2 + v_3}{2} \maps{-U} (-1)^{(n+1)/2}\ket |t> + \frac{v_2 + v_3}{2},
\end{equation}
which gives the required transduction with transduction complexity
\[
\frac{\|v_2 + v_3\|^2}{4} 
\le 
\frac12\sA[\|v_2\|^2 + \|v_3\|^2]
=
\frac12\sC[\sum_{k=1}^n {\sB[C_k + \frac{1}{C_k}]}].
\qedhere
\]
\end{proof}

%\rf{thm:welded} is a special case of \rf{thm:hierarchical}, and the latter gives an $O(n)$ upper bound for even $n$ as well.
An example of this construction for the welded tree graph with depth $3$ is given in \rf{fig:flatted}.
This analysis is identical to the one done in \rf{sec:welded}.
\rf{thm:hierarchical} also gives an $O(n)$ upper bound on the hitting time for welded tree graphs with even depth $n$.
For instance, \rf{fig:short} features the case of even depth $2$.

\def\shortlinegraph#1#2#3#4#5#6#7{
    \begin{scope}
    [
        vertex/.style={circle, draw, inner sep=2pt},
        inA/.style={vertex, fill=yellow!30},
        inB/.style={vertex, fill=green!30},
    ]
    \matrix [column sep=9mm, ampersand replacement=\&]
    {
        \node[vertex, fill=\inputcolor!50] (S') {$s'$}; \&
        \node[inA] (S) {$s_0$}; \&
        \node[inB] (S1) {$s_1$}; \&
        \node[inA] (S2) {$s_2$}; \&
        \node[inB] (S3) {$s_3$}; \&
        \node[inA] (S4) {$s_4$}; \&
        \node[inB] (T) {$s_7$}; \&
        \node[vertex,fill=\markedcolor!50] (T') {$t'$}; \\
    };
    \end{scope}
%    \draw (S') <- (S) -> (S1) <- (S2) -> (S3) <- (S4) -> (S5) <- (S6) -> (T) <- (T');
    \draw[<-] (S') --node[above, blue]{$#1$}node[below]{1} (S);
    \draw[->] (S) --node[above, blue]{$#2$}node[below]{$2$} (S1);
    \draw[<-] (S1) --node[above, blue]{$#3$}node[below]{$4$} (S2);
    \draw[->] (S2) --node[above, blue]{$#4$}node[below]{$8$} (S3);
    \draw[<-] (S3) --node[above, blue]{$#5$}node[below]{$4$} (S4);
    \draw[->] (S4) --node[above, blue]{$#6$}node[below]{$2$} (T);
    \draw[<-] (T) --node[above, blue]{$#7$}node[below]{4} (T');
}

\myfigure{\label{fig:short}}
{
An example for the welded tree graph with even depth $2$.
This time $W_0=1$ and $W_{n+1}=4$.
\\
(a) The vector $\ket |s> + \ket |v_2> - (-1)^{(n+1)/2}\ket |t>$.
(b) The vector $\ket |s> + \ket |v_3> + (-1)^{(n+1)/2}\ket |t>$.
\\
Notations are the same as in \rf{fig:flatted}.
}
{
\[
\begin{tikzpicture}
\shortlinegraph{1}{\sqrt 2}{-1}{-\sqrt 2}{2}{\sqrt 2}{1}
\draw (S') +(-1,0) node {(a)};
\end{tikzpicture}
\]
\[
\begin{tikzpicture}
\shortlinegraph{1}{-\sqrt{\frac12}}{-1}{\sqrt{\frac12}}{\frac 12}{-\sqrt{\frac12}}{-1}
\draw (S') +(-1,0) node {(b)};
\end{tikzpicture}
\]
}

\section*{Acknowledgements}
The author would like to thank Shankar Balasubramanian and Tongyang Li for answering some questions related to~\cite{balasubramanian:hierarchicalGraphs} and for useful comments on an early version of this manuscript.

This research is supported by the Latvian Quantum Initiative under European Union Recovery and Resilience Facility project no. 2.3.1.1.i.0/1/22/I/CFLA/001 and the QuantERA project QOPT.

\bibliographystyle{habbrvM}
{
\small
\bibliography{belov}
}

\end{document}